\documentclass[11pt,showpacs,preprintnumbers,superscriptaddress,amsmath,amssymb,nofootinbib]{revtex4}

\usepackage{color}
\usepackage[dvipdfm]{graphicx}
\usepackage{array}

\newcolumntype{I}{!{\vrule width 1.3pt}}

\begin{document}
\title{Higgs boson decays to $\gamma\gamma$ and $Z \gamma$ in models with Higgs extensions}
\author{Cheng-Wei Chiang}
\affiliation{Department of Physics and Center for Mathematics and Theoretical Physics,
National Central University, Chungli, Taiwan 32001, ROC}
\affiliation{Institute of Physics, Academia Sinica, Taipei, Taiwan 11529, ROC}
\affiliation{Physics Division, National Center for Theoretical Sciences, Hsinchu, Taiwan 30013, ROC}
\author{Kei Yagyu}
\affiliation{Department of Physics and Center for Mathematics and Theoretical Physics,
National Central University, Chungli, Taiwan 32001, ROC}


\begin{abstract}
The decays of a Higgs boson to the $\gamma\gamma$ and $Z\gamma$ final states are purely quantum mechanical phenomena that are closely related to each other.  We study the effects of an extended Higgs sector on the decay rates of the two modes.  We propose that a simultaneous determination of them and the $ZZ$ mode is a useful way to see whether the Higgs boson recently observed by the LHC experiments is of the standard model type or could be a member of a larger Higgs sector.
\end{abstract}

\maketitle

\section{Introduction \label{sec:intro}}

The quest for the origin of elementary particle masses is arguably one of the most important tasks in current high energy physics.  According to the standard model (SM), a scalar field is employed to break the electroweak (EW) symmetry down to the electromagnetic (EM) symmetry, $SU(2)_L \times U(1)_Y \to U(1)_{\rm EM}$, giving masses to the $W$ and $Z$ bosons that mediate weak interactions.  This so-called Higgs mechanism \cite{Higgs} is achieved when the scalar field spontaneously acquires a nonzero vacuum expectation value (VEV) due to the instability in its potential.  As a consequence, the SM predicts the existence of a spin-0 Higgs boson.  With the introduction of Yukawa interactions, fermionic particles can obtain their masses from the same Higgs field as well.  Therefore, the discovery of the Higgs boson does not only complete the particle spectrum of the SM, but also reveals the secrets of EW symmetry breaking and mass.

Recently, both ATLAS and CMS Collaborations \cite{Higgs_search} of the CERN Large Hadron Collider (LHC) have reported the observation of a Higgs-like resonance at around 125 GeV at $\sim 5\sigma$ level through the combination of $ZZ$ and $\gamma\gamma$ channels.  
In particular, in the $\gamma\gamma$ channel, 
the observed cross section is $1.9\pm0.5$ and $1.6\pm 0.4$ times larger than the expected cross section in the SM at the ATLAS and the CMS, respectively.
Measurements in the $WW$, $Wh/Zh$ with $h \to b \bar b$, and the $\tau^+ \tau^-$ channels are in general consistent with the SM expectations. Around this mass region, another decay channel that is closely related to the diphoton mode and clean in the LHC environment is the $Z\gamma$ mode.  At the leading order in the SM, both the $\gamma\gamma$~\cite{hgamgam_SM} and $Z\gamma$~\cite{hZgam_SM} are loop processes mediated by the same particles.  New particles beyond the SM can change their relative magnitudes.  Although the $ZZ$ decay occurs at tree level and is less sensitive to new particle contributions, the rate depends on how the EW symmetry is broken.  Therefore, a simultaneous measurement of their production rates will be helpful in diagnosing the observed Higgs-like particle.

In view of the 125 GeV Higgs boson, denoted by $h$, we consider models that have only an extended Higgs sector for simplicity.  There are some recent studies in the literature about the $h\to\gamma\gamma$ decay in models with an extended Higgs sector~\cite{hgamgam_2HDM,hgamgam_HTM,hgamgam_Zee}.  In this paper, we investigate both the $h\to\gamma\gamma$ and $Z\gamma$ decays in models with Higgs extensions based on various physics motivations.  We assume that $h$ is SM-like, meaning that the couplings of $h$ with fermions ($h\bar{f}f$) as well as the weak gauge bosons ($hVV$) are the same as the SM ones.  
This is consistent with the current experimental observations.  In this case, the production cross section of $h$ is the same as in the SM, and the deviations in the event rates of $\gamma\gamma$ and $Z\gamma$ final states from the SM predictions are purely due to the modified decay rates of the two modes.
We study how the decay rates of $h \to \gamma\gamma$ and $Z \gamma$ are modified (see also \cite{Carena_Low_Wagner}).

This paper is organized as follows.  Section~\ref{sec:models} classifies models with simple extensions in the Higgs sector and give the corresponding quantum numbers for new scalar fields under the SM electroweak group. 
The formulae for the modified decay rates of $h\to \gamma\gamma$ and $h\to Z\gamma$ are also given in this section.  Constraints from perturbativity and vacuum stability on the model parameters are discussed in Section~\ref{sec:results}.  The effects of new scalar bosons on the two decay modes are also analyzed in detail.  Our findings are summarized in Section~\ref{sec:summary}.

\section{Models and modified decay rates \label{sec:models}}
In general, models with an extended Higgs sector often contain charged Higgs bosons that, among other phenomena, can contribute to the $h\to\gamma\gamma$ and $Z\gamma$ decays through the loop effect.  Although there are many possibilities for the extended Higgs sector, we discuss those with extra $SU(2)_L$ singlets $S$ (with $Y=1$ and $Y=2$), doublet $D$ (with $Y=1/2$), and triplets $T$ (with $Y=0$ and $Y=1$), whose charge assignments are given in Table~\ref{TB:charge}.

\begin{table}[t]
{\renewcommand\arraystretch{1.5}
\begin{tabular}{c cc c cc}
\hline\hline
& ~~$S^+$~~ & ~~$S^{++}$~~ & ~~$D$~~ & ~~$T_0$~~ & ~~$T_1$~~ \\
\hline
$SU(2)_L$ & {\bf 1} & {\bf 1} & {\bf 2} & {\bf 3} & {\bf 3}\\
$U(1)_Y$ & +1 & +2 & +1/2  & 0 & +1 \\
\hline\hline
\end{tabular}}
\caption{Charge assignments for extra scalar fields under the $SU(2)_L\times U(1)_Y$ gauge symmetry.
}
\label{TB:charge}
\end{table}
\begin{table}[t]
{\renewcommand\arraystretch{1.5}
\begin{tabular}{ccccccccc ccccc}
\hline\hline
Model    & ~1~ & ~2~ & ~~3~~ & ~~4~~ & ~5~ & ~6~ & ~7~~ & ~~8~~ & ~9~ & ~10~ & ~11~ & ~12~ & ~13~ \\
\hline
$S^+$    & 1 & 0 & 0 & 1 & 0 & 0 & 0 & 2 & 0 & 0  & 1  & 0  & 1  \\
$S^{++}$ & 0 & 0 & 0 & 1 & 0 & 1 & 1 & 0 & 0 & 0  & 0  & 0  & 0  \\
$D$      & 0 & 1 & 0 & 0 & 0 & 1 & 0 & 0 & 2 & 0  & 1  & 1  & 0  \\
$T_0$    & 0 & 0 & 1 & 0 & 0 & 0 & 1 & 0 & 0 & 2  & 0  & 1  & 1  \\
$T_1$    & 0 & 0 & 0 & 0 & 1 & 0 & 0 & 0 & 0 & 0  & 0  & 0  & 0  \\
\hline\hline
\end{tabular}}
\caption{Models considered in this work and their number of extra scalar fields.}
\label{TB:models}
\vspace{-3mm}
\end{table}

We consider three distinct classes of extended Higgs sectors: 
(Class-I) models with one singly-charged scalar boson, (Class-II) those with one singly-charged and one doubly-charged scalar bosons, and (Class-III) those with two singly-charged scalar bosons. 
According to the representations listed in Table~\ref{TB:charge}, there are 
three, four, and six possibilities for Class-I, Class-II and Class-III, respectively, all listed in Table~\ref{TB:models}. 
Examples of models in Class-I (Models~1 to 3) include the two Higgs doublet model~\cite{2HDM} and the minimal supersymmetric SM.  Models in Class-II (Models~4 to 7) include the Higgs triplet model~\cite{HTM}
and Zee-Babu model~\cite{ZeeBabu}.  Finally, models in Class-III (Models~8 to 13) include those 
where tiny Majorana neutrino masses are generated via higher-loop processes~\cite{radiative_seesaw}. 


The modified decay rates of $h\to \gamma\gamma$ and $Z\gamma$ due to the charged scalar boson loops can be expressed in the case where 
the couplings of $h$ to the SM particles are SM-like by
\begin{align}
\hspace{-5mm}\Gamma_{\gamma\gamma}&=
\frac{\sqrt{2}G_F\alpha_{\text{em}}^2m_h^3}{256\pi^3 }
 \left|\sum_{i}Q_{S_i}^2\frac{\lambda_{SSh}^{ii}}{v}I_{S}^i+\sum_{f}Q_f^2N_c^fI_{f}+I_W\right|^2, \\
\Gamma_{Z\gamma}&=\frac{\sqrt{2}G_F\alpha_{\text{em}}^2m_h^3}{128\pi^3}\left(1-\frac{m_Z^2}{m_h^2}\right)^3
\left|\sum_{i,j} Q_{S_i}g_{SSZ}^{ij}\frac{\lambda_{SSh}^{ij}}{v}J_S^{ij}+\sum_f Q_fN_c^fJ_{f}+J_{W}\right|^2 ~,
\end{align}
where $G_F$ is the Fermi decay constant, $v=1/(\sqrt{2}G_F)^{1/2}$ is the Higgs VEV,  
$m_h$ is the Higgs boson mass, $m_Z$ is the $Z$ boson mass, $Q_X$ is the electric charge of particle $X$, $N_c^f$ is the color factor of the fermion $f$.  The loop functions for the scalar contribution $I_S^i$ and $J_S^{ij}$ are given by
\begin{align}
I_S^i&=\frac{2v^2}{m_h^2}[1+2m_{S_i}^2C_0(0,0,m_h^2,m_{S_i},m_{S_i},m_{S_i})] ~, \\
J_S^{ij}&=\frac{2v^2}{e(m_h^2-m_Z^2)}\Bigg\{1+[m_{S_i}^2C_0(0,m_Z^2,m_h^2,m_{S_i},m_{S_i},m_{S_j})+(i\leftrightarrow j) ]\notag\\
&+\frac{m_Z^2}{e(m_h^2-m_Z^2)}[B_0(m_h^2,m_{S_i},m_{S_j})-B_0(m_Z^2,m_{S_i},m_{S_j})]\Bigg\}
 ~, 
\end{align}
in terms of the Passarino-Veltman functions $B_0$ and $C_0$ defined in Ref.~\cite{PV}, where $m_{S_i}$ is the mass of the charged scalar boson $S_i$.  
The loop functions for the $W$ boson ($I_W$ and $J_W$) as well as the fermion $f$ ($I_f$ and $J_f$) contributions are given in Ref.~\cite{Djouadi}.  
We note that the value of the $C_0$ function asymptotically approaches $-1/(2m_{S_i}^2)$ when $m_{S_i}$ is much larger than $m_h$ or $m_Z$.  Therefore, as long as $\lambda_{SSh}^{ij}$ is taken to be a fixed value, deviations in $\Gamma_{\gamma\gamma}$ and $\Gamma_{Z\gamma}$ 
from those in the SM vanish in the limit of $m_{S_i}\to \infty$. 

The couplings between the charged scalar bosons $S_i$ and the $Z$ boson as well as $h$ are defined by 
\begin{align}
\label{eq:Slagrangian}
\mathcal{L}_S&=-\lambda_{SSh}^{ij}S_iS_j^*h
+ig_{SSZ}^{ij}(\partial_\mu S_i S_j^*+S_i\partial_\mu S_j^*) Z^\mu +\text{h.c.}
\end{align}

In models of Class-I, the coupling constants in Eq.~(\ref{eq:Slagrangian}) are degenerate and given by
\begin{align}
g_{SSZ}^{ij}=g_{SSZ}=\frac{g}{c_W}(I_3-s_W^2Q_{S}),\quad \lambda_{SSh}^{ij}=\lambda_{SSh}=\frac{2}{v}(m_{S^+}^2-M^2) ~,\label{hss1}
\end{align}
where $M^2$ is the coefficient of the quadratic term of the additional scalar field that is unrelated to the Higgs VEV, $I_3$ is the third isospin component of the singly-charged scalar boson $S^\pm$, and $s_W=\sin\theta_W$, $c_W=\cos\theta_W$ with $\theta_W$ being the weak mixing angle.

In models of Class-II, the couplings $g_{SSZ}^{ij}$ and $\lambda_{SSh}^{ij}$ are proportional to $\delta^{ij}$, associated with the singly-charged scalar boson ($i=1$) and the doubly-charged scalar boson ($i=2$).  These couplings are given as 
\begin{align}
g_{SSZ}^{ii} &= \frac{g}{c_W}(I_3^i-s_W^2Q_{S_i}),\\
\lambda_{SSh}^{11}&=\frac{2}{v}(m_{S^+}^2-M_+^2),\quad \lambda_{SSh}^{22}=\frac{2}{v}(m_{S^{++}}^2-M_{++}^2), 
\end{align}
where $M_+$ and $M_{++}$ have the same dimension as $M$ given in Eq.~(\ref{hss1}) and are generally independent parameters\footnote{In Model~5, $M_+$ and $M_{++}$ are the same, as they derive from the same multiplet.}.
 
In Class-III models, on the other hand, the two singly-charged charged scalar bosons $S_1^\pm$ and $S_2^\pm$ generally mix with each other, so that the expressions for $g_{SSZ}^{ij}$ and $\lambda_{SSh}^{ij}$ ($i,j=1,2$) can be quite different from those given in Eq.~(\ref{hss1}).  In this case, the coupling $g_{SSZ}^{ij}$ are written in the mass eigenbasis of the two charged scalar bosons ($S_1^\pm,S_2^\pm$) as
\begin{align}
g_{SSZ}^{ij}=\frac{g}{c_W}\left[
\begin{array}{cc}
I_3^{\varphi_1}c_\theta^2+I_3^{\varphi_2}s_\theta^2-s_W^2 & (I_3^{\varphi_2}-I_3^{\varphi_1})s_{\theta}c_\theta\\
(I_3^{\varphi_2}-I_3^{\varphi_1})s_{\theta}c_\theta & I_3^{\varphi_1}s_\theta^2+I_3^{\varphi_2}c_\theta^2-s_W^2
\end{array}\right], 
\end{align}
where $\theta$ is the mixing angle ($c_\theta=\cos\theta$, $s_\theta=\sin\theta$) connecting between the weak eigenstates ($\varphi_1^\pm$,$\varphi_2^\pm$) and the mass eigenstates:
\begin{align}
\left(
\begin{array}{c}
\varphi_1^\pm \\
\varphi_2^\pm
\end{array}\right)
=\left(
\begin{array}{cc}
c_\theta & -s_\theta \\
s_\theta & c_\theta
\end{array}\right)
\left(
\begin{array}{c}
S_1^\pm \\
S_2^\pm
\end{array}\right), 
\end{align}
with $I_3^{\varphi_{1,2}}$ being the third isospin component of $\varphi_{1,2}^\pm$ and $I_3^{\varphi_2}\geq I_3^{\varphi_1}$. 
%
Note that if $I_3^{\varphi_1}\neq I_3^{\varphi_2}$, the off-diagonal couplings $g_{SSZ}^{12,21}$ are nonzero and can contribute to the $h\to Z\gamma$ decay as well.  $\lambda_{SSh}^{ij}$ can be calculated for Models~8, 9 and 10 as
\begin{align}
\lambda_{SSh}^{11,~22}&=\frac{2}{v}(m_{S_{1,2}^+}^2-M_{1,2}^2c_\theta^2-M_{2,1}^2s_\theta^2\mp M_3^2s_{2\theta}) ~,\\
\lambda_{SSh}^{12}&=\frac{1}{v}\left[(M_1^2-M_2^2)s_{2\theta}-2M_3^2c_{2\theta}\right] ~.
\end{align}
Those for Models~11 and 12 are
\begin{align}
\lambda_{SSh}^{11,~22}&=\frac{2}{v}\left[\left(1-\frac{s_{2\theta}^2}{4}\right)m_{S_{1,2}^+}^2+\frac{s_{2\theta}^2}{4}m_{S_{2,1}^+}^2
-M_{1,2}^2c_\theta^2-M_{2,1}^2s_\theta^2
\right] ~,\\
\lambda_{SSh}^{12}&=\frac{s_{2\theta}}{v}\left[\frac{c_{2\theta}}{2}(m_{S_2^+}^2-m_{S_1^+}^2)+M_1^2-M_2^2\right] ~.
\end{align}
For Model~13,
\begin{align}
\lambda_{SSh}^{ii}=\frac{2}{v}(m_{S_i^+}^2-M_i^2) ~,~~
\lambda_{SSh}^{12}=0 ~~\text{with}~~i=1,2~.
\label{lambda13}
\end{align}
In the above expressions for $\lambda_{SSh}^{ij}$, the dimensionful parameters $M_{1,2,3}$ show up in the scalar potential
\begin{align}
V \supset +M_1^2|\varphi_1|^2+M_2^2|\varphi_2|^2+M_3^2(\varphi_1^\dagger \varphi_2+\text{h.c.}),
\end{align}
where $\varphi_1$ and $\varphi_2$ are the scalar fields including $\varphi_1^\pm$ and $\varphi_2^\pm$, respectively. 
In Models~11 and 12, the parameter corresponding to $M_3$ is absent, while there is another dimensionful parameter $\mu$ defined in the terms $\mu \Phi^\dagger \varphi_1\varphi_2+\text{h.c.}$ that induce mixing between $\varphi_1^\pm$ and $\varphi_2^\pm$, where $\Phi$ is the Higgs doublet field associated with $h$.  In Model~13, there are no parameters corresponding to $M_3$ and $\mu$ and, therefore, there is no mixing at tree level.

We note in passing that the coupling formulae for $\lambda_{SSh}^{ij}$ and $g_{SSZ}^{ij}$ can change if $h$ mixes with the other neutral scalar states and/or when the other scalar fields get nonzero VEV's. 
In such cases, the production cross section of $h$ can also be different from that in the SM. 


%
\begin{table}[t]
\begin{tabular}{ccccc}
\hline\hline 
 $\Gamma_{\text{tot}}$ (MeV)~ & ~${\cal B}_{\gamma\gamma}$ (\%)~ & ~${\cal B}_{Z\gamma}$ (\%)~ 
 & ~${\cal B}_{ZZ}$ (\%)~ & ~$R$ \\
 \hline
 $3.7$ & $0.28$ & $0.18$ & $2.3$ & $0.63$  \\\hline\hline  
\end{tabular}
\caption{
Total decay rate and branching fractions of the Higgs boson $h$ in the SM. }
\label{TB:SM}
\end{table}

\section{Numerical results \label{sec:results}}
To see the correlation between the decay rates of $h\to Z\gamma$ and $\gamma\gamma$, we further define the ratio of 
the two decay rates: 
\begin{align}
R\equiv \Gamma_{Z\gamma} / \Gamma_{\gamma\gamma}.  \label{R_def}
\end{align}
First, we give the SM expectations of the two diboson decays of the Higgs boson in Table~\ref{TB:SM}, where $m_h = 125$ GeV and $m_t=173$ GeV are used.  Next, we show numerical results for the case with an extended Higgs sector.  As mentioned before, we assume that the observed Higgs boson is SM-like in couplings with the gauge bosons and fermions in our numerical studies.  Moreover, we present the results for the case where the charged scalar bosons are at least 300 GeV in mass.

For meaningful discussions, we calculate parameter bounds by considering perturbativity and vacuum stability constraints.  The perturbativity condition requires that the magnitudes of all dimensionless coupling constants do not exceed $2\sqrt{\pi}$. 
For vacuum stability, we require that the scalar potential is bounded from below in the parameter space where the quartic terms dominate.  Combining the two bounds, we obtain the following conditions for Classes~I, II and III, respectively: 
\begin{align}
&-\frac{m_h\pi^{1/4}}{v}<\frac{m_{S^+}^2-M^2}{v^2}<\sqrt{\pi} ~,\text{ for Class-I models}, 
\label{VS1} \\
&-\frac{m_h\pi^{1/4}}{v}<\frac{m_{S^{+,++}}^2-M_{+,++}^2}{v^2}<\sqrt{\pi} ~,\text{ for Class-II models}, 
\label{VS2} \\
&-\frac{m_h\pi^{1/4}}{v}<\frac{m_{S_{1,2}}^2c_\theta^2+m_{S_{2,1}}^2s_\theta^2-M_{1,2}^2}{v^2}<\sqrt{\pi} ~,\text{ for Class-III models}.
\label{VS3}
\end{align}
%
It is noted that these conditions can be modified by some quartic couplings in the scalar potential that are neglected in our analysis.  In the following analysis, we use these conditions to constrain the $M_i$ parameters for a given value of $m_{S_i}$.

The deviations from the SM predictions for the $h\to \gamma\gamma$ and $Z\gamma$ branching fractions can be parametrized as
\begin{align}
\Delta {\cal B}_{\gamma\gamma(Z\gamma)} &= 
\left[
{\cal B}_{\gamma\gamma(Z\gamma)}^{\text{NP}}
-{\cal B}_{\gamma\gamma(Z\gamma)}^{\text{SM}}
\right]
/ {\cal B}_{\gamma\gamma(Z\gamma)}^{\text{SM}} ~,
\end{align}
where ${\cal B}_{\gamma\gamma}^{\text{NP}}$ (${\cal B}_{Z\gamma}^{\text{NP}}$) is the branching fraction of $h\to\gamma\gamma$ ($h\to Z\gamma$) in a Higgs-extended model, while ${\cal B}_{\gamma\gamma}^{\text{SM}}$ (${\cal B}_{Z\gamma}^{\text{SM}}$) is that in the SM.

\begin{figure}[t]
\includegraphics[width=100mm]{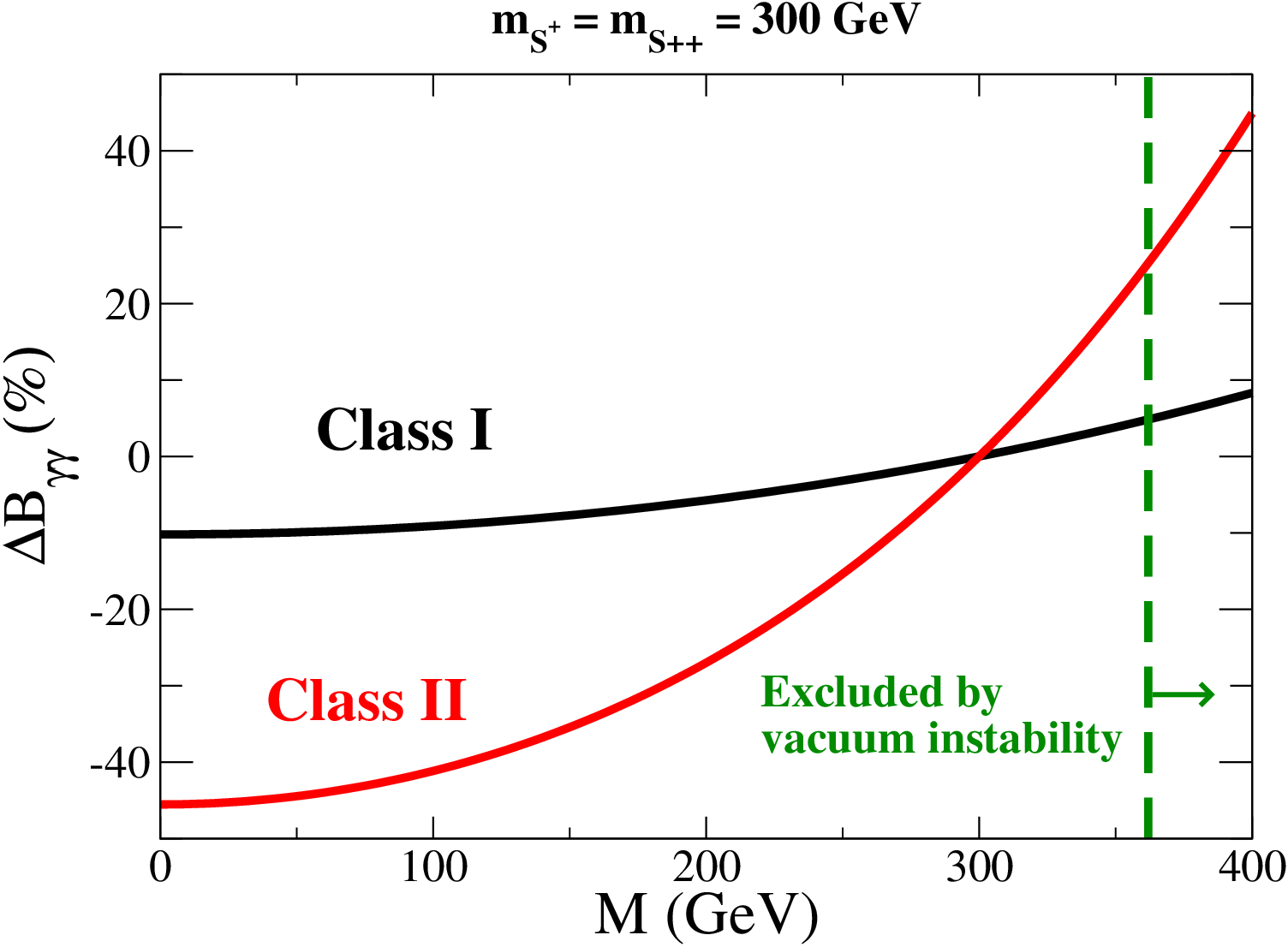}
\caption{Deviation in the branching fraction of $h\to \gamma\gamma$ as a function of $M$.  We take $m_{S^+}=300$ GeV in models of Class~I and $M_+=M_{++}=M$ and $m_{S^+}=m_{S^{++}}=300$ GeV in models of Class~II.  The vertical dashed line indicates the upper limit of $M$ by the vacuum stability condition.
}
\label{fig1}
\end{figure}

In Fig.~\ref{fig1}, the deviation in the branching fraction of the $h\to \gamma\gamma$ mode is shown as a function of $M$ in the case of $m_{S^+}=300$ GeV for Class-I models, and $m_{S^+}=m_{S^{++}}=300$ GeV with $M_+=M_{++}=M$ for Class~II models.  The parameter $M$ is constrained to be $0<M\lesssim 362$ GeV by Eqs.~(\ref{VS1}) and (\ref{VS2}).  For a fixed value of $M$, the value of $\Delta \mathcal{B}_{\gamma\gamma}$ is the same among the models within the same class.  Moreover, the value increases with $M$.  The maximally allowed value of $\Delta \mathcal{B}_{\gamma\gamma}$ is about +4.8\% (+25\%) for Class-I models (Class-II models) when $M$ is about 362 GeV.

\begin{figure}[t]
\includegraphics[width=80mm]{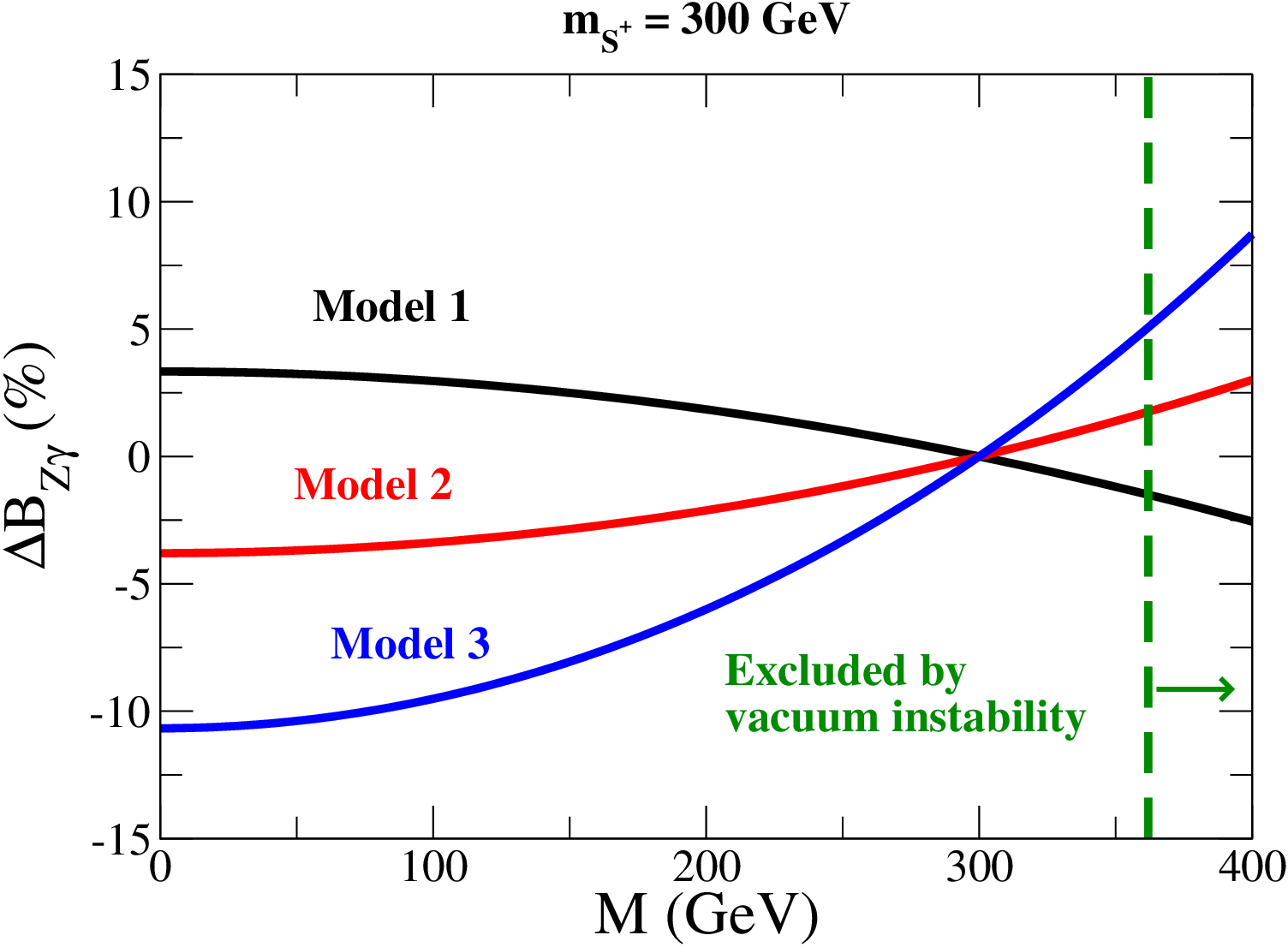}\hspace{3mm}
\includegraphics[width=80mm]{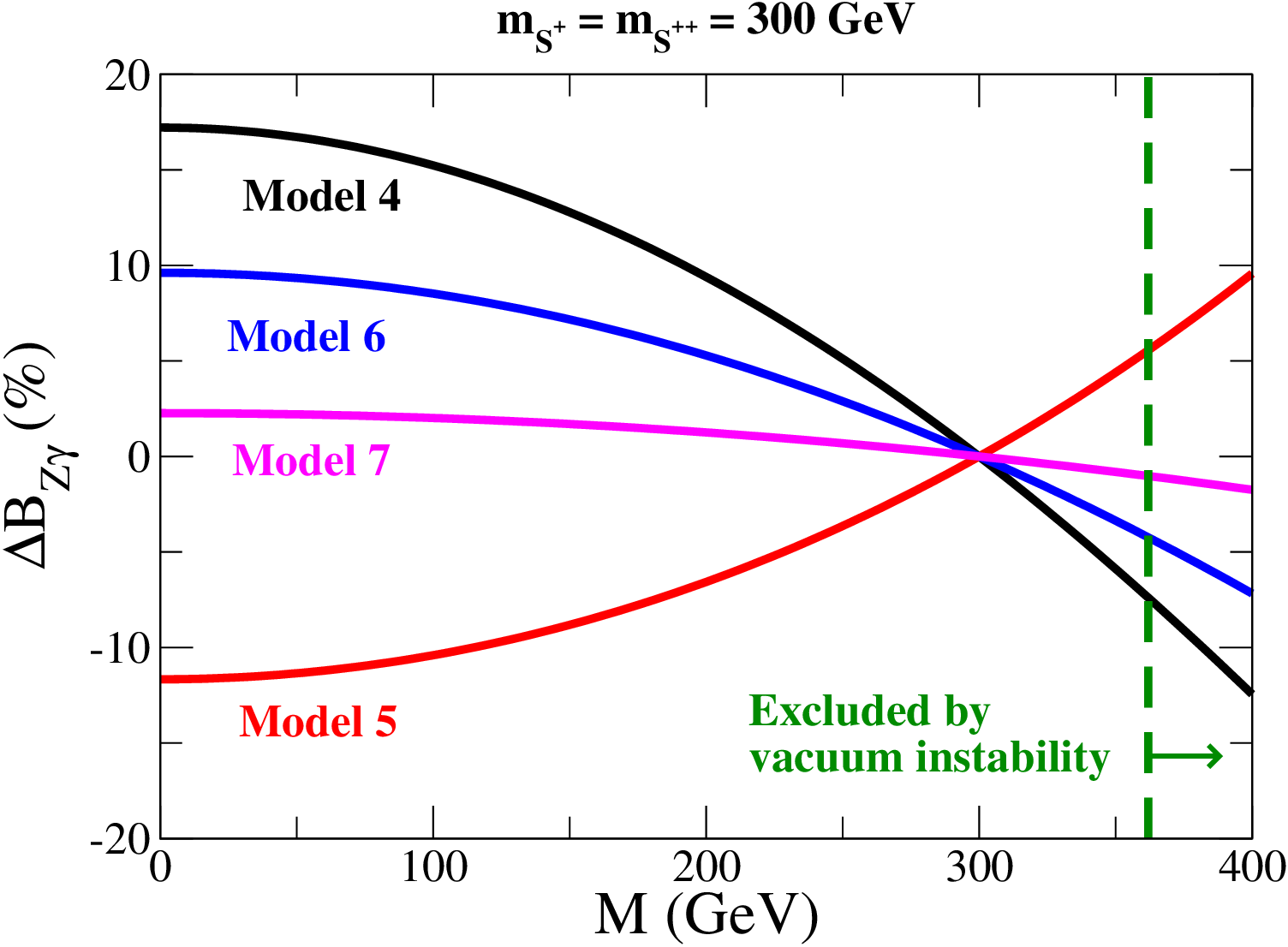}
\caption{ 
Deviation in the branching fraction of $h\to Z\gamma$ as a function of $M$.  We take $m_{S^+}=300$ GeV in models of Class~I and $M_+=M_{++}=M$ and $m_{S^+}=m_{S^{++}}=300$ GeV in models of Class~II.  The vertical dashed line indicates the upper limit of $M$ by the vacuum stability condition.  The left (right) panel shows the results for Class-I models (Class-II models). 
}
\label{fig2}
\end{figure}

\begin{figure}[t]
\begin{center}
\includegraphics[width=80mm]{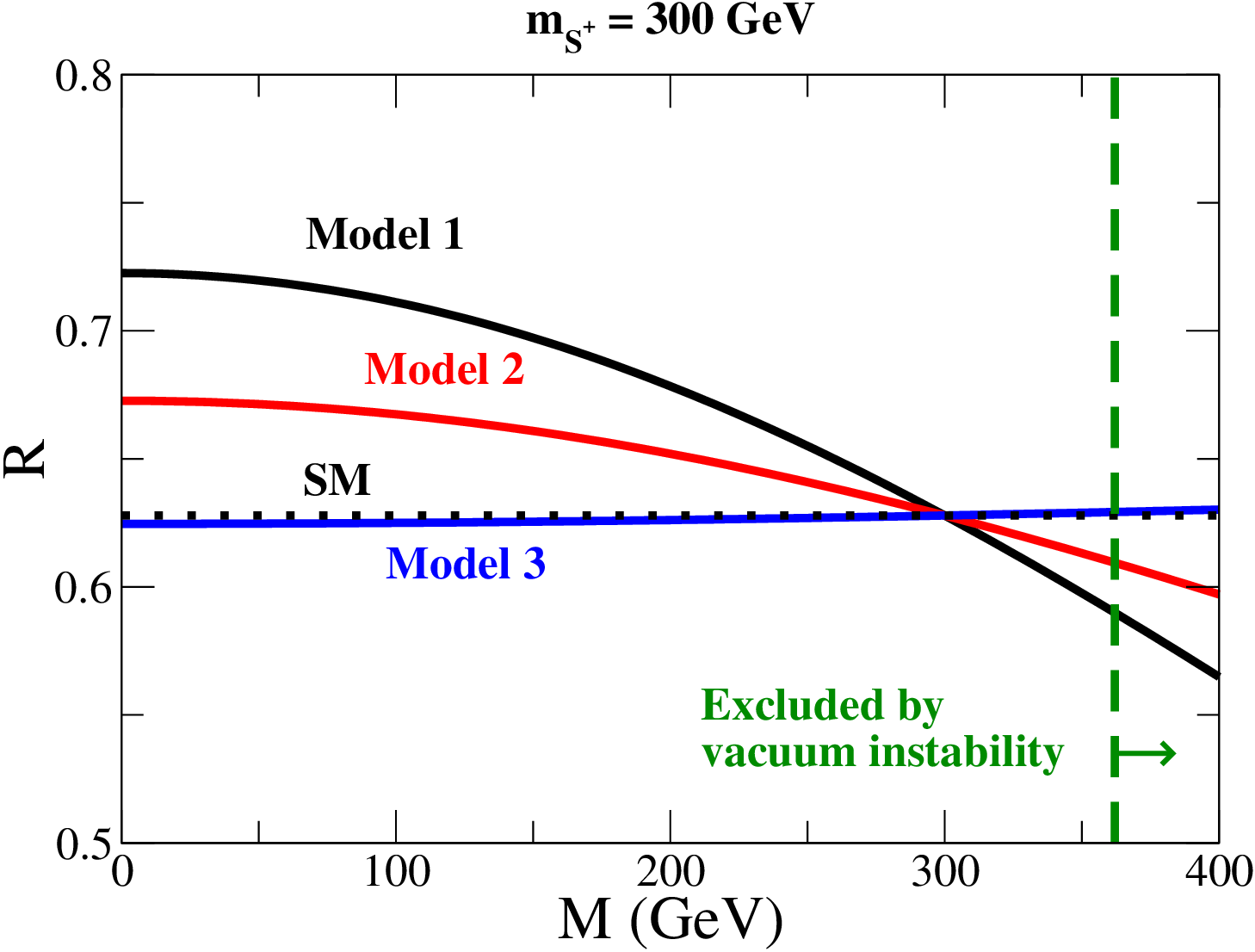}\hspace{3mm}
\includegraphics[width=80mm]{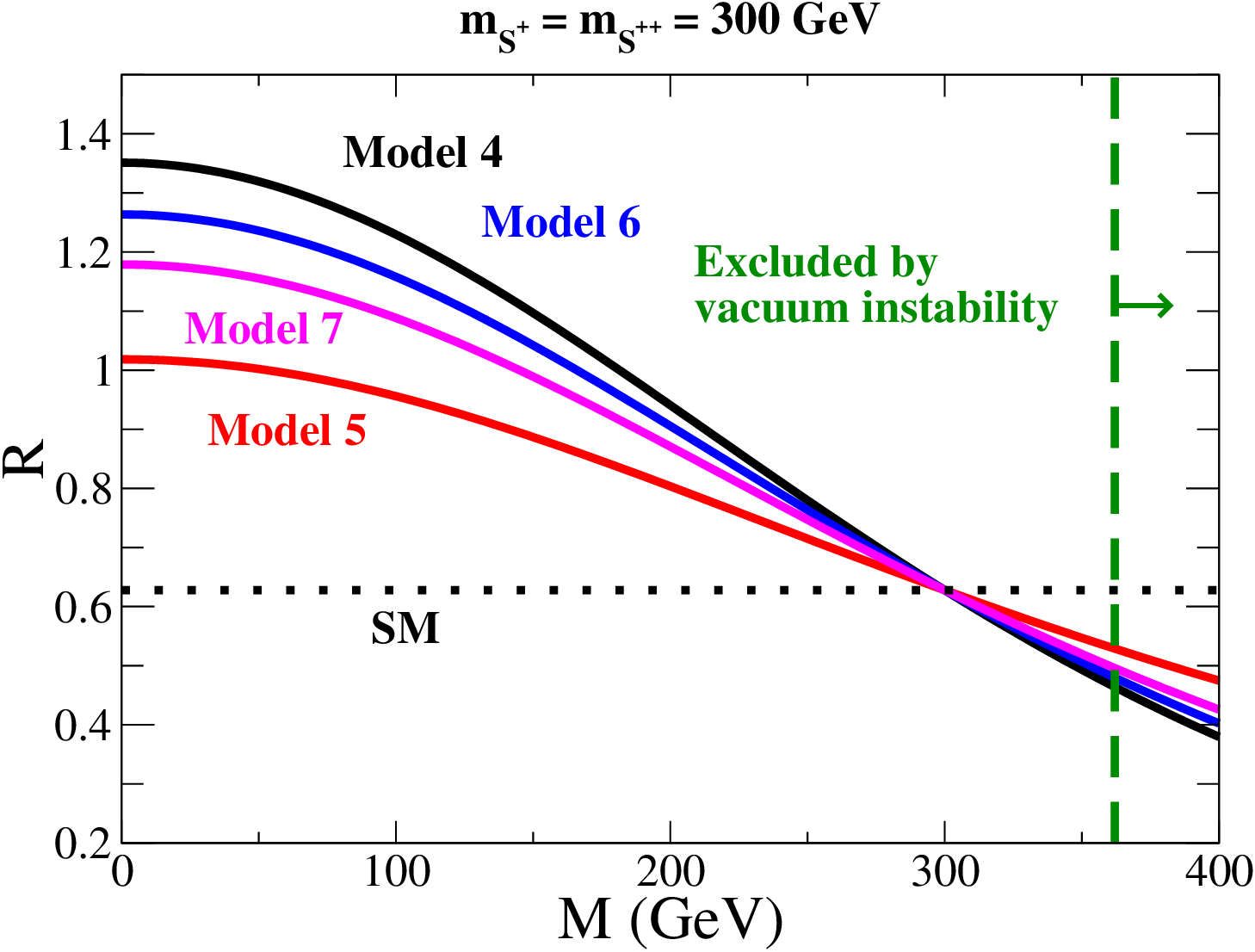}
\caption{The value of $R$ as a function of $M$.  We take $m_{S^+}=300$ GeV in models of Class~I and $M_+=M_{++}=M$ and $m_{S^+}=m_{S^{++}}=300$ GeV in models of Class~II.  The vertical dashed line indicates the upper limit of $M$ by the vacuum stability condition.  The left (right) panel shows the results for Class-I models (Class-II models). 
}
\label{fig3}
\end{center}
\end{figure}

In Fig.~\ref{fig2}, the deviation in the branching fraction of the $h\to Z\gamma$ mode is plotted as a function of $M$ with the same parameter choice as in Fig.~\ref{fig1}.  The value of $\Delta \mathcal{B}_{Z\gamma}$ varies among the models within the same class.  It is seen that models with fields of larger isospin representations tend to have a larger value of $\Delta \mathcal{B}_{Z\gamma}$.  The value of $R$ defined in Eq.~(\ref{R_def}) for models of Class~I and Class~II are shown in Fig.~\ref{fig3}.

For models in Class-III, we consider as an example the case where $m_{S_1^+}=300$ GeV and $m_{S_2^+}=400$ GeV.  We assume that the three dimensionful parameters are the same: $M_1=M_2=M_3=M$.  The maximally allowed value of $M$ depends on the mixing angle $\theta$, but
the strictest upper bound on $M$ from Eq.~(\ref{VS3}) is found when $\theta=0$.  In this case, the upper bound is 362 GeV, which is used in the following numerical analysis.

\begin{figure}[t]
\begin{center}
\includegraphics[width=100mm]{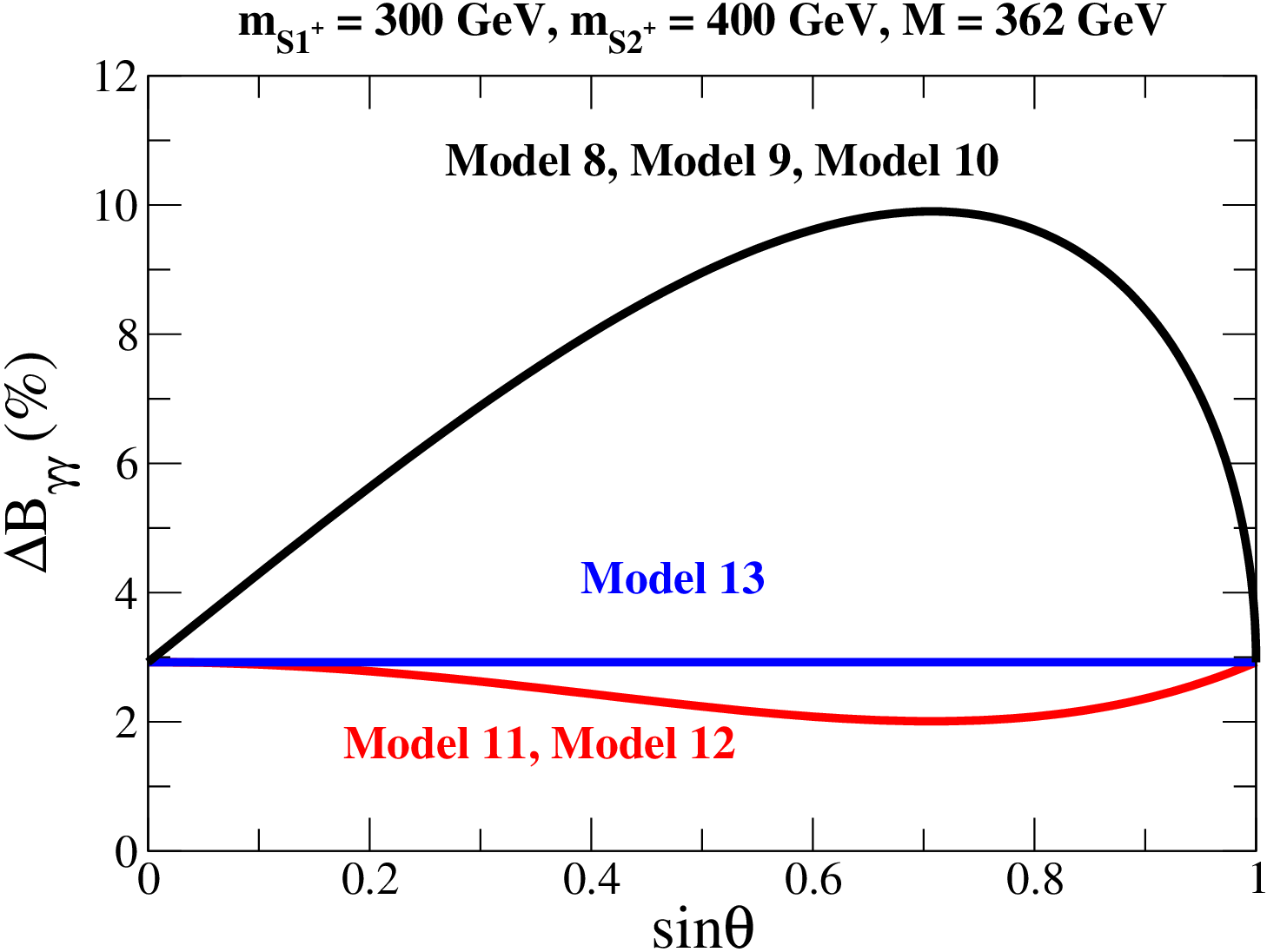}
\caption{Deviation in the branching fraction of $h\to \gamma\gamma$ as a function of $\sin\theta$ for the case with $M_1=M_2=M_3=M=362$ GeV, $m_{S_1^+}=300$ GeV and $m_{S_2^+}=400$ GeV. }
\label{fig4}
\end{center}
\end{figure}

In Fig.~\ref{fig4}, the deviation in the branching fraction of the $h\to \gamma\gamma$ mode is plotted against $\sin\theta$.  All the models in this class (Models 8-13) have positive corrections with the above parameter choice given.  In Models 8-10 (in Models 11-12), the deviation reaches its maximum (minimum) when the mixing is maximal ($\sin\theta \simeq 0.7$).  On the other hand, there is no $\sin\theta$ dependence in Model 13.  The predicted value of $\Delta \mathcal{B}_{\gamma\gamma}$ is the same among Models 8-10 and between Model 11 and Model 12.

\begin{figure}[t]
\begin{center}
\includegraphics[width=80mm]{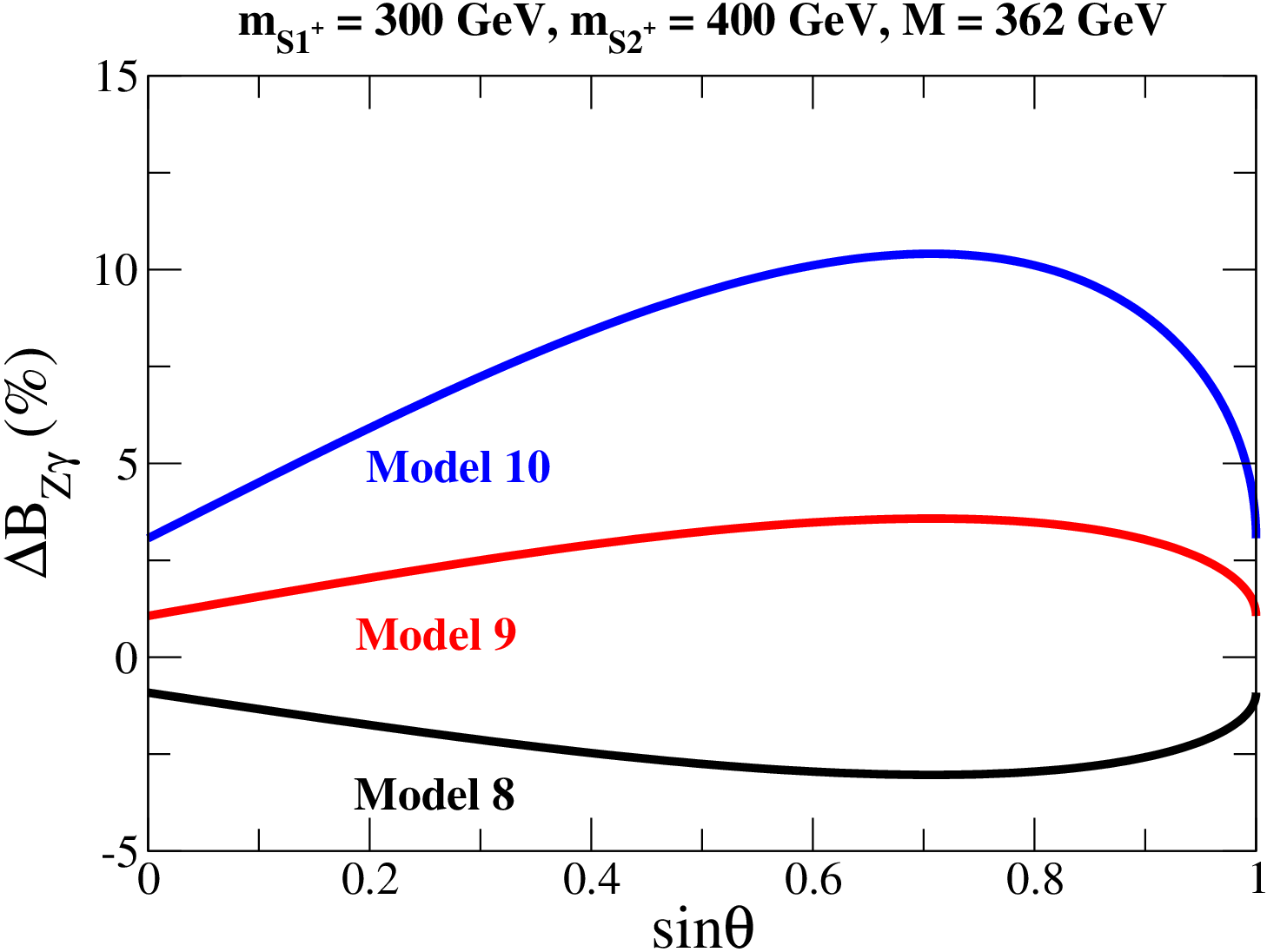}\hspace{3mm}
\includegraphics[width=80mm]{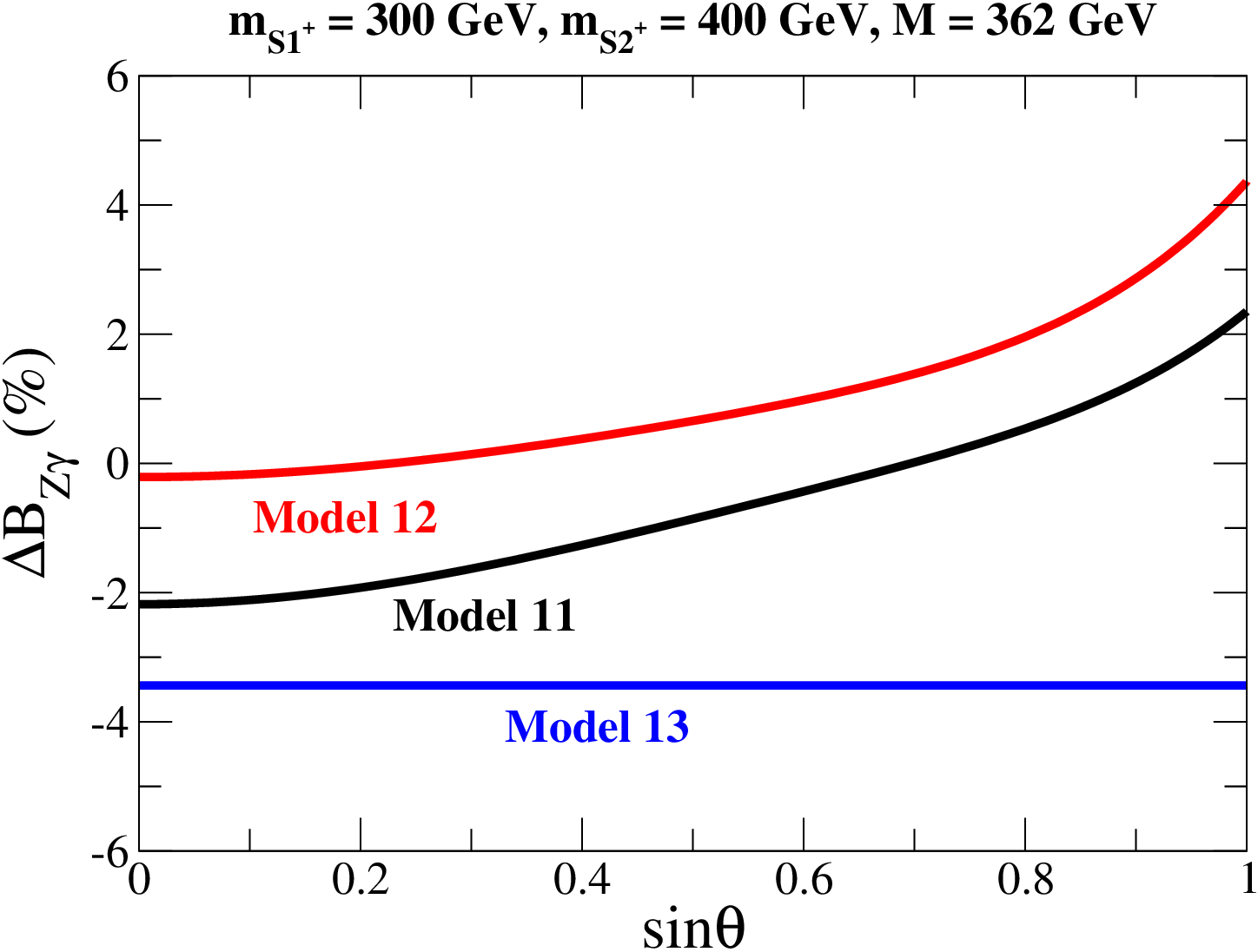}
\caption{Deviation in the branching fraction of $h\to Z\gamma$ as a function of $\sin\theta$ for the case with $M_1=M_2=M_3=M=362$ GeV, $m_{S_1^+}=300$ GeV and $m_{S_2^+}=400$ GeV. 
The left (right) panel shows the results for Models~8, 9 and 10 (Models~11, 12 and 13). }
\label{fig5}
\end{center}
\end{figure}

\begin{figure}[t]
\vspace{5mm}
\begin{center}
\includegraphics[width=70mm]{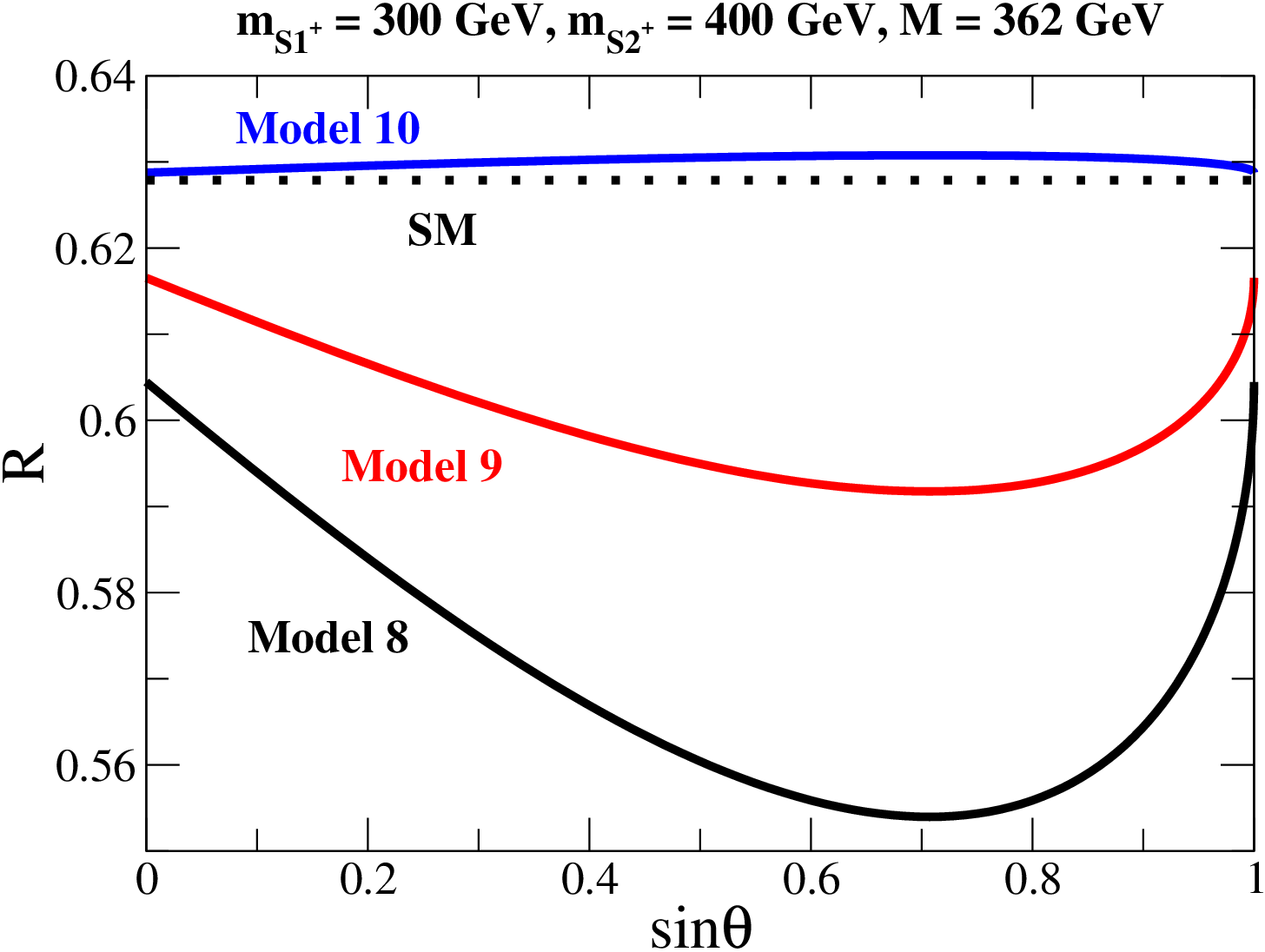}\hspace{3mm}
\includegraphics[width=70mm]{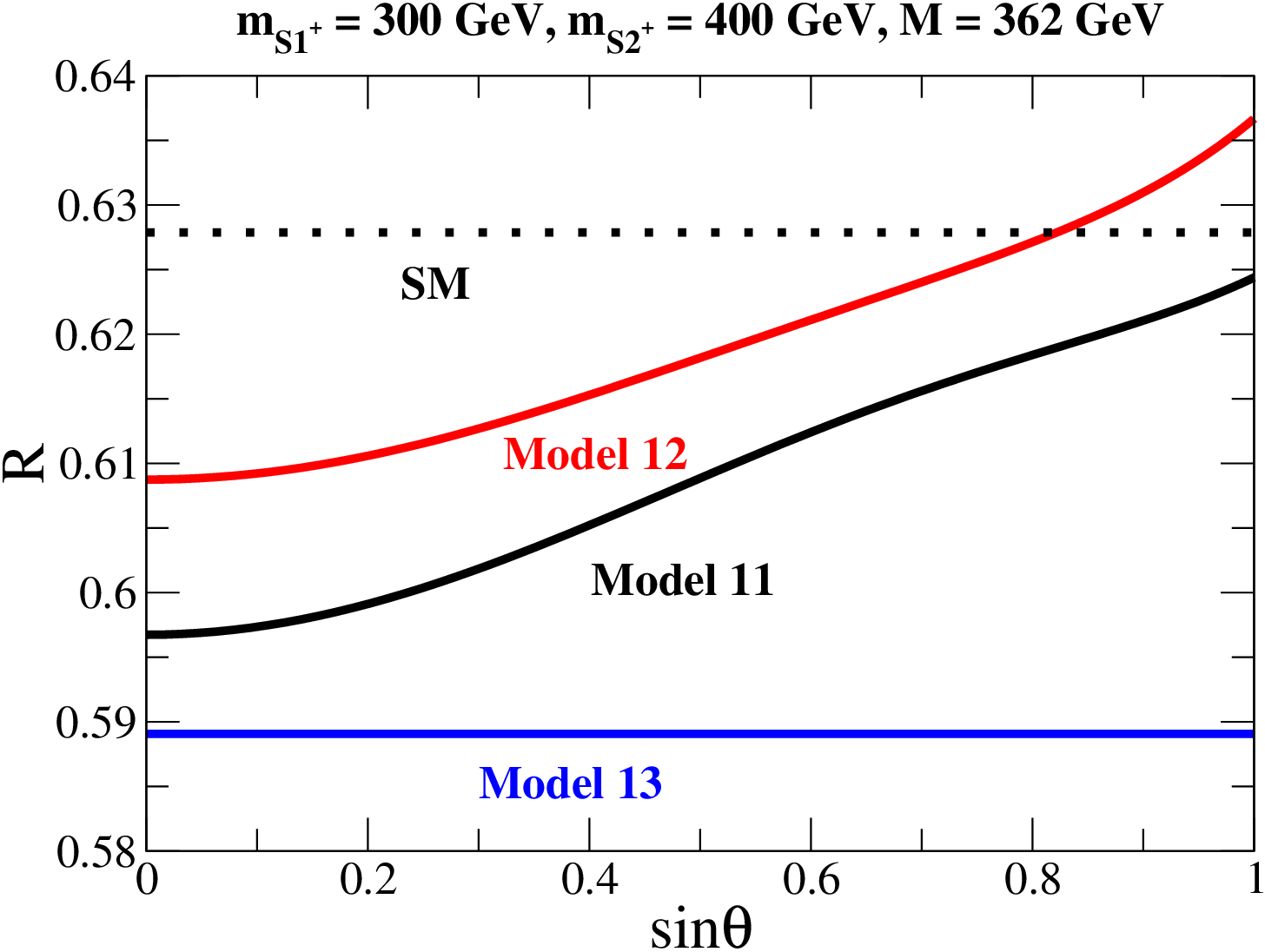}
\caption{The ratio $R$ defined in Eq.~(\ref{R_def}) as a function of $\sin\theta$ for the case with $M_1=M_2=M_3=M=362$ GeV, $m_{S_1^+}=300$ GeV and $m_{S_2^+}=400$ GeV. 
The SM prediction is indicated by the dotted line for comparison. 
The left (right) panel shows the result in Models~8, 9 and 10 (Models~11, 12 and 13). }
\label{fig6}
\end{center}
\vspace{-5mm}
\end{figure}

In Fig.~\ref{fig5}, the deviation in the branching fraction of the $h\to Z\gamma$ mode is shown in models of Class~III as a function of $\sin\theta$.  The left panel shows the results in Models~8, 9 and 10, while the right panel shows those in Models~11, 12 and 13.  There are differences in the values of $\Delta \mathcal{B}_{Z\gamma}$ among Models 8-10 and Models 11-12.  As seen in Fig.~\ref{fig2}, the model with larger isospin representation fields tends to get a larger value of $\Delta \mathcal{B}_{Z\gamma}$.  
In Fig.~\ref{fig6}, we show the ratio $R$ in models Class~III in contrast with the values for the SM.

\begin{figure}[t]
\begin{center}
\includegraphics[width=75mm]{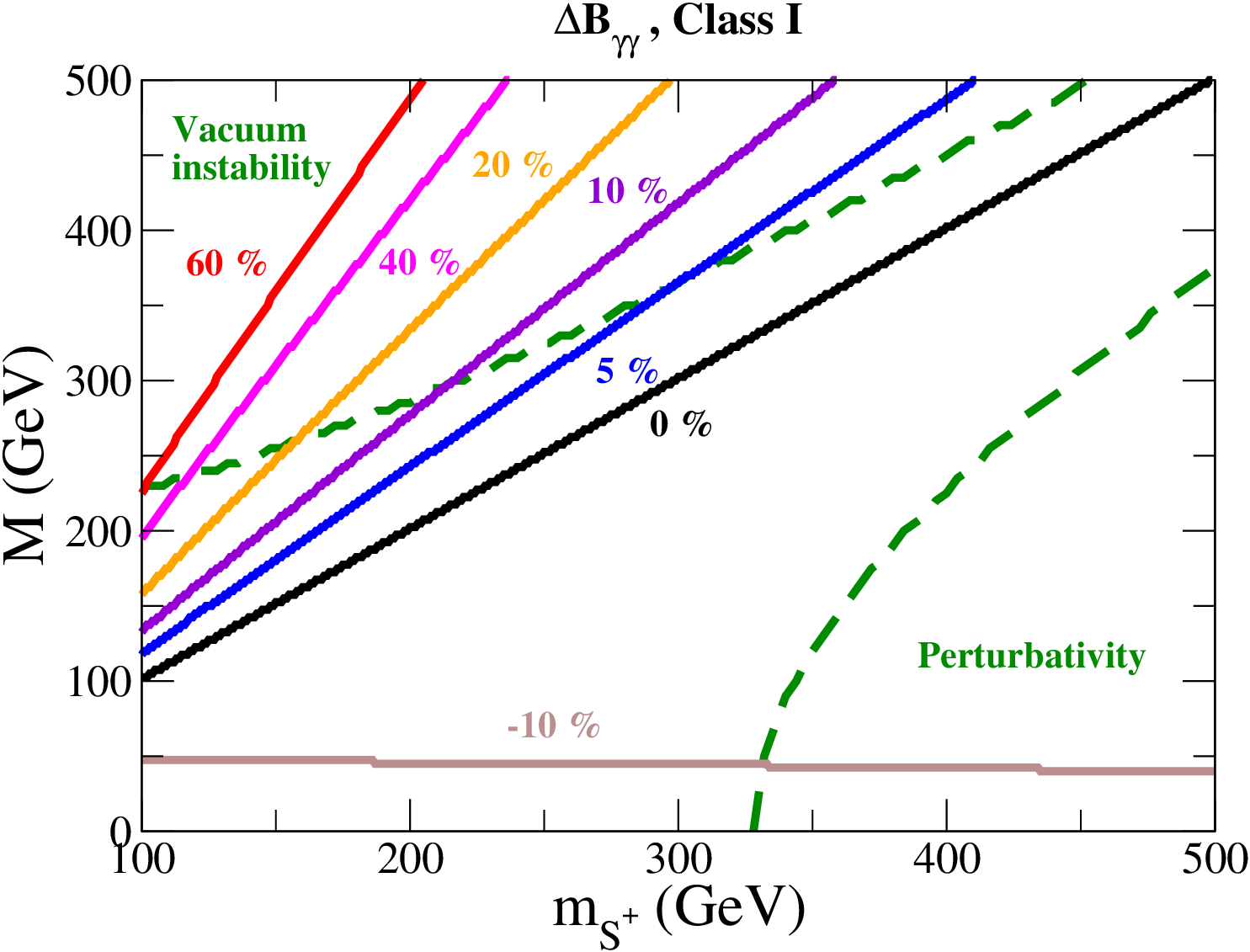}\hspace{4mm}
\includegraphics[width=75mm]{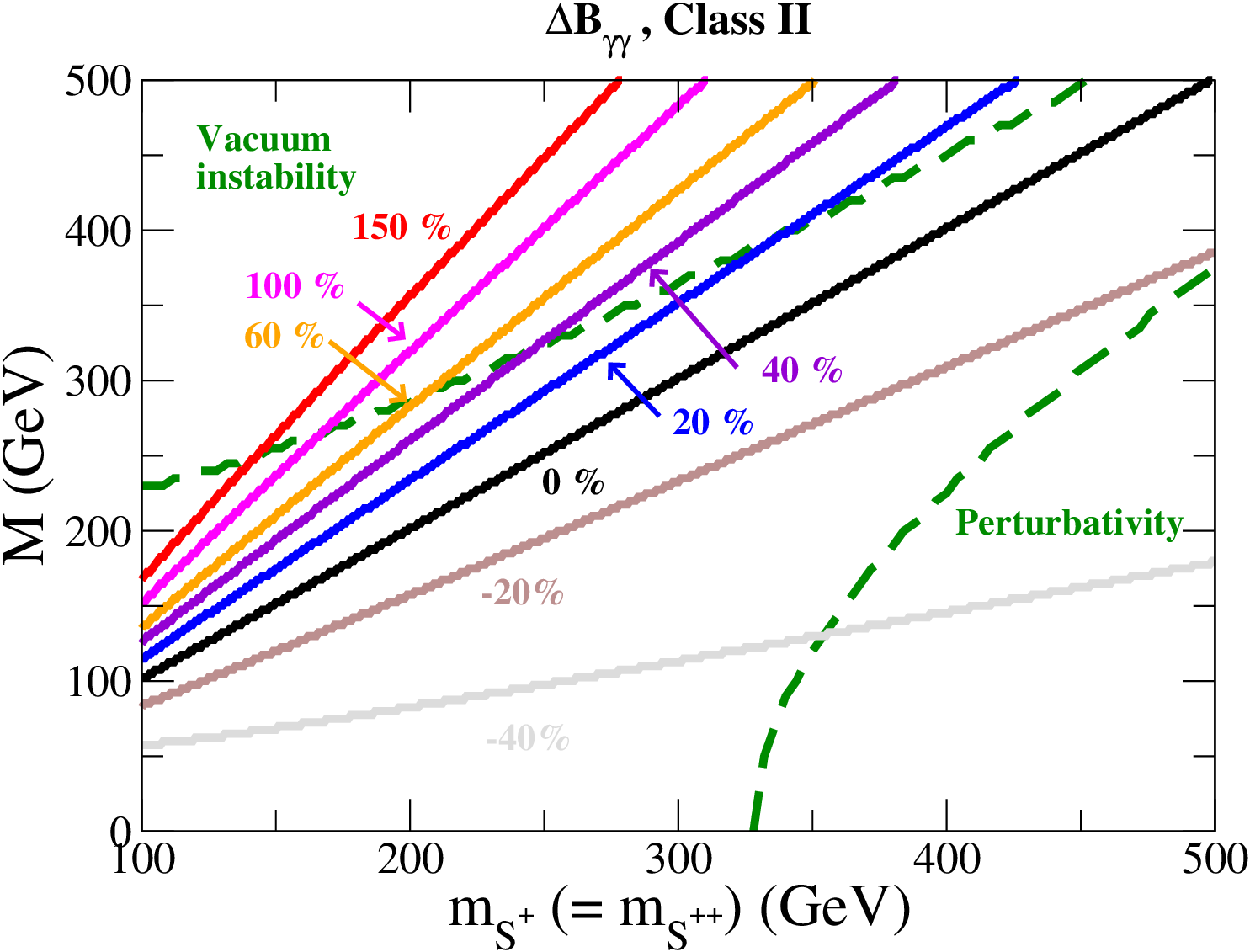}
\caption{Contour plots of $\Delta \mathcal{B}_{\gamma\gamma}$ in the $m_{S^+}$-$M$ plane in models of Class I (left panel) and Class II (right panel). 
We take $m_{S^+}=m_{S^{++}}$ and $M_+=M_{++}=M$ in Class-II models. 
The dashed curves indicate parameter regions excluded by the constraints of vacuum stability and perturbativity. 
}
\label{fig7}
\end{center}
\end{figure}

Finally, we show the contour plots of $\Delta \mathcal{B}_{\gamma\gamma}$ in the $m_{S^+}$-$M$ plane in Fig.~\ref{fig7}.  The left (right) panel shows the results in modes of Class I (Class II).  We take $m_{S^+}=m_{S^{++}}$ and $M_+=M_{++}=M$ in models of Class II. 
As indicated by the dashed curves in both plots, the upper left corner of the parameter space is excluded by the vacuum stability and the lower right corner by the parturbativity. 
In models of Class I, it is impossible to get a deviation of more than +60\% for $\Delta \mathcal{B}_{\gamma\gamma}$ as long as the mass of the charged scalar boson $m_{S^+}$ is greater than 100 GeV because of the constraint from vacuum stability. 
When $m_{S^+}$ is greater than 200 GeV, $\Delta \mathcal{B}_{\gamma\gamma}$ is less than +10\%. 
In comparison, for models of Class II, deviations of more than +60\% are possible for $\Delta \mathcal{B}_{\gamma\gamma}$ when the charged scalar boson masses are smaller than 200 GeV.  Therefore, Class-II models can better explain the current observation of excess production in the diphoton channel at the LHC.

\section{Summary \label{sec:summary}}

With the observation of a Higgs boson $h$ of mass $125$ GeV by ATLAS and CMS, it would be interesting to diagnose whether $h$ is standard model-like (SM-like) or part of a larger Higgs sector.  We thus consider and classify models with simple Higgs extensions.  We have imposed the perturbativity and vacuum stability constraints on the model parameters.
We have studied the neutral diboson decays of $h$, assuming that it has SM-like couplings with the weak bosons and fermions.  In our framework, the $ZZ$ mode is virtually unaffected, whereas the $\gamma\gamma$ and $Z\gamma$ modes can be modified by a few to a few tens of percent.  A simultaneous determination of their branching fractions is thus useful in exploring the possibility of an extended Higgs sector.

\smallskip
{\it Acknowledgments }
The authors thank C.~M.~Kuo for useful discussions.  This research was supported in part by the National Science Council of Taiwan, R.~O.~C.
under Grant Nos.~NSC-100-2628-M-008-003-MY4 and NSC-101-2811-M-008-014-.


\end{document}